\documentclass[prl,aps,twocolumn,floats,showpacspsfig]{revtex4}
\usepackage{amssymb}

\usepackage{epsfig}

\newcommand{\be}{\begin{equation}}
\newcommand{\ee}{\end{equation}}
\newcommand{\bea}{\begin{eqnarray}}
\newcommand{\eea}{\end{eqnarray}}

\newcommand{\p}{\partial}
\newcommand{\s}{\sigma}

\newcommand{\la}{\langle}
\newcommand{\ra}{\rangle}

\newcommand{\ri}{\mbox{i}}

\newcommand{\dw}{\downarrow}
\newcommand{\up}{\uparrow}

\begin{document}
\title{Self-energy of a nodal fermion in a d-wave superconductor.}

\author{ A. V. Chubukov$^{1}$ and A.  M. Tsvelik$^{2}$}
\affiliation{$^{1}$ Department of Physics, 
 University of Wisconsin, Madison, WI53706}
\affiliation{$^{2}$ Department of  Condensed Matter Physics and Materials Science, Brookhaven National Laboratory, Upton, NY 11973-5000, USA}
\date{\today}

\begin{abstract}
We re-consider the self-energy of a nodal (Dirac) 
fermion in a 2D $d-$wave 
 superconductor.  A conventional belief is that 
Im $\Sigma (\omega, T) \sim {\text max} (\omega^3, T^3)$.
We show that  $\Sigma (\omega, k, T)$ for $k$ along the nodal direction
 is actually a complex function of $\omega, T$, and the deviation from the
 mass shell. In particular,  the second-order self-energy 
 diverges at a finite $T$ when either $\omega$ or $k-k_F$ vanish.
We show that  the full summation of infinite 
diagrammatic series recovers
a finite result for $\Sigma$, but the full ARPES spectral function 
  is non-monotonic and has a kink whose location compared to the mass shell 
  differs qualitatively for spin-and charge-mediated interactions. 
  \end{abstract}

\maketitle

The physics of nodal quasiparticles in a 2D $d-$wave superconductor attracted a considerable attention since the early days of high $T_c$ era~\cite{exp_nodal,valla,campuz}
because of its
 universality~\cite{lee} and a relation to field theoretical 
studies of Dirac fermions~\cite{alyosha}.
In the normal state of the cuprates,
 ARPES and other studies have found that  the imaginary part of the 
 fermionic self-energy  along the nodal direction,  $Im \Sigma (\omega, T)$,
 is roughly linear in both  frequency
 and temperature~\cite{valla,campuz,varma}. In the superconducting state, 
a generic belief is
 that $Im \Sigma (\omega, T)$ should become smaller 
 as whatever the mechanism is for the quasiparticle scattering
 in the normal state, it must be weakened in the superconducting state
 due to  the gap opening~\cite{peter}.
 Recent ARPES results do show indeed that once bilayer
 splitting is resolved, the measured 
$Im \Sigma (\omega, T)$ decreases below $T_c$ faster than 
 linear~\cite{boris,kaminsk,valla_2}. 

In an $s-$wave superconductor, $Im \Sigma (\omega, T)$ 
 at $\omega < 3\Delta$ vanishes  at $T=0$ and is 
 exponentially small at finite $T$~\cite{scalap}. 
In a $d_{x^2-y^2}-$ superconductor, the gap 
vanishes along the diagonal directions
 in the Brillouin zone. Hence
 the  scattering into low-energy states near the nodes
  gives rise  to  power-law $\omega$ and $T$ dependences of $Im \Sigma (\omega, T)$  at the lowest $\omega$ and $T$.

Several groups have previously computed~\cite{titov,quinlan,duffy} 
 the quasiparticle lifetime  of a nodal fermion in a 
BCS $d-$wave superconductor at $T=0$ and at $k =k_F$, 
and found the cubic frequency 
dependence  $Im \Sigma (\omega, T=0) \propto \omega^3$.
 A common belief was  that (i) this dependence survives at 
 arbitrary ratio of $\omega$ and $\epsilon_k = v_F (k-k_F)$ along the nodal direction, and (ii) for a clean system, 
the temperature dependence is the same as  the frequency dependence, 
 i.e.,  $Im \Sigma (\omega, T)$ scales as  $\omega^3$ or 
$T^3$, wichever is larger. 

In the present communication, we dispute this common belief. 
We show that the perturbative self-energy for a nodal fermion
 actually scales as $T^{7/2}/\sqrt{\omega}$ at $T >> \omega$ and generic
 $\omega/\epsilon_k$  and logarithmically  
 diverges at the mass shell, $\omega = \epsilon_k$. 
This singular behavior has been largely missed in earlier 
 studies  (with the exception of \cite{khvesh}, 
where $T^{7/2}/\sqrt{\omega}$ has been reported).
 We show that the origin of this singular behavior
 is the same as of the mass-shell singularity in 
a 2D Landau Fermi liquid (LFL)~\cite{metzner,maslov}. Like in a LFL, 
 the singular behavior of the self-energy  for a Dirac fermion is 
eliminated once infinite series of divergent terms are summed up,  
 but the resulting spectral function has a kink at some finite $\omega - \epsilon_k$. We found that the kink is located 
 at $0> \omega > \epsilon_k$, or at $0>\epsilon_k > \omega $ depending
 on whether  the effective interaction is in the spin or in the 
charge channel. 
From this perspective, detailed
 ARPES studies of nodal spectral function
 can  qualitatively distinguish between 
 the theories for the cuprates  based on spin or charge fluctuations. 

We consider a model of a $d-$wave superconductor with 
a quasiparticle dispersion 
$E_k = \sqrt{\epsilon^2_k + \Delta^2_k} = |{\bf k}|$,
 where ${\bf k} = (v_F k_{\perp},  v_\Delta k_{\parallel})$,
 and $k_\perp$ and $k_{\parallel}$ are  
  deviations from the nodal point transverse to  
and along the Fermi surface , respectively ($\epsilon_k = v_F k_{\perp}$).
 The  fermion-fermion interaction has both charge and spin components,
 $U_c (q)$ and $U_s (q)$. 
 We explicitly verified in lengthy calculations that only 
intra-nodal interactions, governed by $U_c (0)$ and $U_s (0)$ contribute 
 to  the self-energy  along the nodal direction, and we consider only these terms. Inter-nodal interactions do contribute
 to the self-energy away from the nodal direction, which we did not consider.
 
We first re-analyzed the self-energy to the second order in the perturbation. 
We carried out calculations both in the Nambu formalism and
 using normal and anomalous Green's functions, and obtained
 identical results in both approaches. 
In the Nambu formalism, the   mean field action near a given pair or 
nodes at ${\bf k}_F$ and $-{\bf k}_F$ 
is described by the four-component massless Dirac fermions (Nambu spinors) and
  has the following Lorentz invariant form: 
\bea
{\cal L} = \ri\bar\psi(\gamma_0\p_{\tau} + v_F\gamma_1\p_x + v_{\Delta}\gamma_2\p_y)\psi
\label{4}
\eea
where $\psi^{+}(k) = \left(\psi^+_{\dw} (k), \psi_{\up} (-k),\psi^+_{\up} (k), \psi_{\dw} (-k)\right)$, $\bar\psi = \psi^+\gamma_0$, $\gamma_0 = I\otimes\s^y, \gamma_1 = I\otimes\s^x, \gamma_2 =\s^z\otimes\s^z$ and 
$\s^a$ are the Pauli matrices. The free-fermion Green's function is $G = \la\psi \bar\psi\ra = \ri\gamma_{\mu}p_{\mu}/p^2$. The fermion-fermion interaction is
 the sum of the bilinear products of charge density and spin density operators
$\psi_{\alpha}^{\dagger} \psi_{\alpha}$ and $\psi_{\alpha}^{\dagger} \sigma_{\alpha \beta} \psi_{\beta}$, respectively.  The perturbation theory is constructed in the same way as in LFL. 

In the conventional formalism, which we will follow below, the self-energy 
 is obtained by summing up contributions from normal and anomalous Green's functions of intermediate fermions. The 
 expression for $Im \Sigma (\omega, k).  
T)$ obtained in this formalism is presented in~\cite{peter,duffy}. 
We verified and confirmed their result. The authors of ~\cite{peter} 
 analyzed the self-energy numerically and argued that it follows $max 
(\omega^3, T^3)$ behavior. We evaluated the self-energy analytically
  and found new, singular  behavior at $\omega \approx \epsilon_k$ at $T=0$, and  at $\omega, \epsilon_k \ll T$ at a finite $T$.  
To understand the origin of the singularity
 and to establish connection to the earlier work on the mass-shell singularity in  LFL, it is convenient to view the self-energy as a  
convolution of the fermionic and  particle-hole propagators:
\begin{eqnarray}
&&Im \Sigma (\omega, k, T) \propto U^2 \int d^2 q d \Omega
\left(\coth{\frac{\Omega}{2T}} - \tanh{\frac{\omega + \Omega}{2T}}\right) \nonumber \\
&& Im \chi_0 (\Omega, q) Im G_0 (\omega + \Omega, k +q)
\label{11}
\end{eqnarray}
where $G_0$ is the BCS fermionic Green function. 
 In  non-Nambu  notations,  the polarization bubble is the sum of normal and anomalous components, and is given by 
\begin{equation}
\chi_0 (\Omega, q) \propto \int d^2 {\bf l} \frac{E_+ + E_-}{E_+ E_-}~\frac{\frac{q^2}{4} -l^2 + E_+ E_-}{(E_+ + E_-)^2 - (\Omega + i 0)^2}
\label{12}
\end{equation}
where $E_{\pm} = |{\bf l} \pm {\bf q}/2|$. 
At small $\Omega << q$, $Im \chi_0 (\Omega, q) \propto \Omega$,
as has already been established before~\cite{titov}. 
 The convolution of this form 
 with the linear frequency dependence of the fermionic density
 of states in a $d-$wave superconductor
 gives rise to $Im \Sigma (\omega, \epsilon_k, T) 
\propto (\omega^3)$ at $\omega \sim T$ and generic $\epsilon_k/\omega$, 
 in agreement with ~\cite{peter}.  
The behavior of the self-energy near the mass shell is, however, determined by $|\Omega| \approx q$, where 
 $\chi_0 (\Omega, q)$ is singular. The
 singularity originates  from the integration in (\ref{12}) 
over $l < q/2$, and over directions of 
${\bf l}$ which are nearly parallel to ${\bf q}$.
In this range, $E_{+} + E_{-} \approx q$, i.e., the denominator almost vanishes
 {\it in a finite range of internal $l$}, while the
  numerator in (\ref{12}) remains finite.  Expanding 
$E_{+}  +  E_{-}$ in the angle $\theta$ between ${\bf l}$ and ${\bf q}$ as 
  $E_+ + E_- \approx q[ 1 + l^2 \theta^2/(q^2 - 4 l^2) + ...]$, substituting into (\ref{12}) and 
 integrating over $\theta$,  we obtain
\begin{equation}
\chi_0 (\Omega, q) = A \frac{q^2}{\sqrt{q^2 - (\Omega +i \delta)^2}}
\label{14}  
\end{equation}
where $A >0$. We see that 
 $Im \chi_0 (\Omega, q)$ is nonzero when 
$~~~|\Omega| >q$, and it diverges at $|\Omega| \rightarrow q$.  

 Substituting
 $Im \chi_0 (\Omega, q)$ and $Im G_0 (\omega + \Omega, k +q) 
\propto |\omega + \Omega|~ \delta ((\omega + \Omega)^2 - ({\bf k} + {\bf q})^2)$ into (\ref{11}), we find
 that on the mass shell (i.e., when at $\omega = k$),  
 the position of the branch cut singularity in $Im \chi (\Omega, q)$ 
 and the location of the $\delta-$function in $Im G (\omega + \Omega, k +q)$ coincide provided  ${\bf k}$ and ${\bf q}$ are either parallel (for $\omega'>0$)
 or antiparallel (when $\omega'<0$). Integrating near these directions, we obtain the singular part of the self-energy. At $T=0$,
we find that $Im \Sigma (\omega, \epsilon_k, T=0)$ is non-zero only when 
$\omega < \epsilon_k$ (in conventional variables where 
$E_k = |\epsilon_k|$, and for $\omega <0$, relevant to ARPES experiments), 
and is discontinuous at $\omega =\epsilon_k$ 
($Re \Sigma$ diverges logarithmically at this point). Explicitly,
\begin{eqnarray}
&&Im \Sigma (\omega \approx \epsilon_k, T=0) = \nonumber \\
&&\frac{2 \pi^3}{105}~(u^2_c + 3 u^2_s)~
 \left(\frac{v_F}{v_\Delta}\right)^2  \frac{\omega^3}{E^2_F}~ 
\theta \left (\omega - |\epsilon_k|\right)
\label{15_1}
\end{eqnarray}
where $E_F = v_F k_F/2$  and $u_{c,s} = m U_{c,s} (0)/(2\pi)$.

At a finite $T$, we obtained in three  limits:
\begin{eqnarray*}
Im \Sigma (\omega =0, k, T) \sim  &\frac{T^{7/2}}{\sqrt{|\epsilon_k|}},&
 |\epsilon_k| \ll T , \nonumber \\
& T \epsilon^2_k e^{-|\epsilon_k|/2T},& |\epsilon_k|  \gg T.  \nonumber \\
Im \Sigma (\omega, k=0, T) \sim &
\frac{T^{7/2}}{\sqrt{|\omega|}},& |\omega| \ll T , \nonumber \\
&|\omega|^3 \left (1 + O\left(\frac{T}{|\omega|} \right)^{3/2}\right),&~~~|\omega| >> T.
\end{eqnarray*}
\begin{eqnarray}
&&Im \Sigma (\omega \approx \epsilon_k, T) \sim |\omega|^3 
\theta (\epsilon_k -\omega) + \nonumber \\
&& \frac{T^{3/2}}{ \sqrt{|\omega|}} (a T^2 + b \omega^2)
 \log{|\frac{\epsilon_k}{\epsilon_k -\omega}|}
\end{eqnarray} 
where $a, b = O(1)$. We see that at finite $T$, the second-order self-energy 
diverges logarithmically on the mass shell, and also diverges as 
$1/\sqrt{|\omega|}$ 
when both $\omega$ and $\epsilon_k$ tend to zero~\cite{khvesh}. 
This result is very different from a simple $T^3$ behavior. 
The latter is only recovered when $|\omega| 
\sim T$, and the system is at some distance away from the mass shell. 

As  we have said, the singularity in the self-energy for a Dirac fermion is 
 similar to the mass-shell singularity in the fermionic self-energy a  
 2D LFL~\cite{metzner,maslov}.
 In both cases, the singularity originates from the fact that the
 the pole in the fermionic Green's function almost coincides with the branch-cut  in the polarization bubble~\cite{maslov}. 
There exists, however, an important distinction between the two cases, which makes the singularity for Dirac fermions stronger than in a Fermi liquid. In a liquid,
 the quasiparticle dispersion changes sign at the Fermi surface, i.e., $\epsilon_{k_F +q} = q \cos \theta$ can be both positive and negative.
Accordingly,  
$Im \Sigma (\omega, T) \propto \int  d^2 q d \theta d \omega^\prime~q~ 
\delta (\omega^\prime + (\omega - k) - q \cos \theta)/\sqrt{(\omega^\prime)^2 - q^2}$.
  and the angular integration reduces the $\delta$ function to another square-root, so that  the subsequent momentum integral only logarithmically depends on the 
$\omega -k$, i.e., on the distance from the mass shell. For Dirac fermions, the
 dispersion $E_k = |k|$ is {\it positive}, and $E_{k_F +q} \propto |q| >0$. 
The angle integral still accounts for the logarithmic singularity at 
$\omega =E_k$, but in addition,  the combination of the $\delta$ function from the fermionic propagator and the square-root singularity in the polarization operator gives rise to the extra  $(T/|\omega|)^{1/2}$ factor in the self-energy.

The discontinuity of the self-energy on the mass shell at $T=0$ and the
 divergence at $T >0$ imply that higher-order diagrams for the self-energy 
 may also be relevant. Evaluating higher-order self-energy diagrams
with extra particle-hole bubbles,
 we find that they
actually diverge at the mass shell already at $T=0$, 
and the divergences proliferate with the
 order of perturbation. Like in a LFL, the most divergent diagrams 
 form ladder (RPA) series~\cite{maslov},
 such that the full self-energy can be still 
represented by Eq. (\ref{11}), but now $\chi_0 (\Omega, q)$ has to be replaced by the full  susceptibility $\chi (\Omega, q)$.  This computational procedure is justified when $u_{c,s} \ll 1$.

\begin{figure}[ht]
\begin{center}
\epsfxsize=0.47\textwidth
\epsfbox{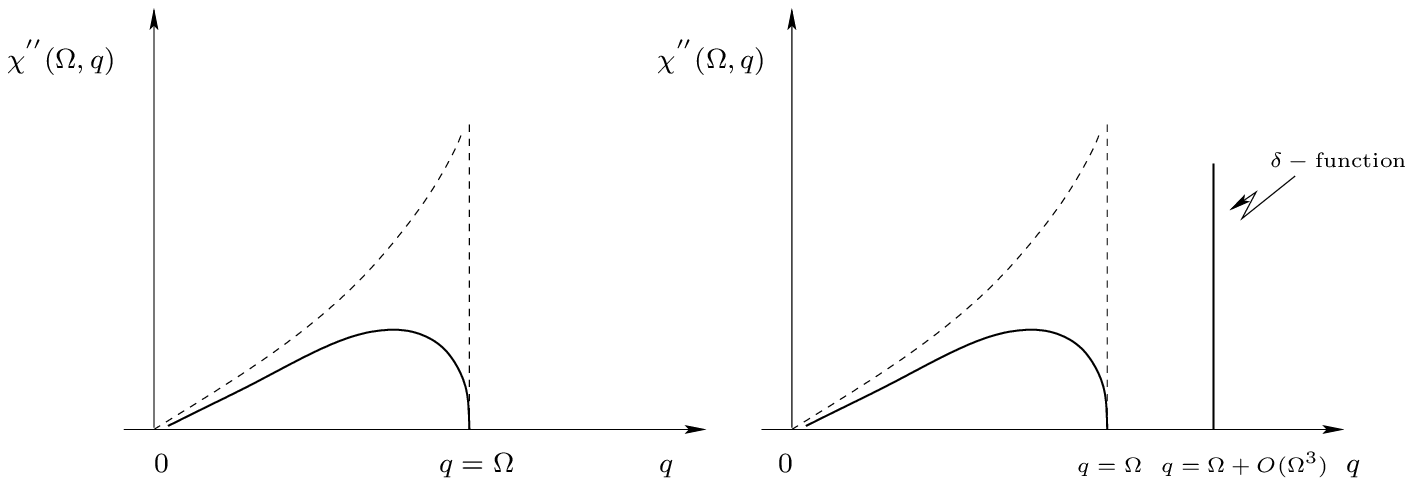}
\end{center}
\caption{The behavior of the imaginary part of the 
full bosonic propagator Im $\chi (\Omega, q)$ for interacting Dirac
 fermions. Panels a and b are for the interaction in the charge
 and  in spin channel, respectively. 
 For spin interaction, $\chi (\Omega, q)$ develops a zero-sound pole
 outside the particle-hole continuum. The dashed line is the imaginary part of the bosonic propagator for free Dirac fermions.}
\label{fig1}
\end{figure}

At this
 stage, it becomes essential whether the effective interaction is in the charge or in the spin channel, as the RPA renormalizations in charge and spin channels 
have opposite signs. Summing up the bubble and the ladder diagrams, we
 obtain the following expressions for the full $\chi_c (\Omega, q)$ and $\chi_s (\Omega, q)$
\begin{equation}
\chi_c (\Omega, q) = \frac{\chi_0 (\Omega, q)}{1 + U_c \chi_0 (\Omega, q)}~~
\chi_s (\Omega, q) = \frac{\chi_0 (\Omega, q)}{1 - U_s \chi_0 (\Omega, q)}~~ 
\label{aa}
\end{equation}
where $\chi_0 (\Omega, q)$ is given by (\ref{14}) (recall that 
$U_{c,s} >0$ and $Re \chi_0 (\Omega, q) >0$). 
 For a charge-mediated interaction, there is no pole outside the
 particle-hole continuum. Accordingly, the RPA renormalization only
 eliminates the square-root singularity in $Im \chi (\Omega, q)$ at
 $\Omega =q$. The full $Im \chi (\Omega, q)$ 
 then behaves as in Fig. \ref{fig1}a: it initially follows $Im \chi_0 (\Omega, q)$ 
and increases at approaching  $\Omega =q$, but passes 
 through a maximum and vanishes at $\Omega =q$. For $\chi_s (\Omega, q)$, the
 sign of the renormalization is different, and the interaction not
 only eliminates the  square-root divergence at $\Omega =q$, but also 
 generates
 a zero-sound-type pole outside the particle-hole continuum, at $\Omega = q (1 - B u^2_s q^2)$, $B \sim 1/k^{2}_F > 0$, as on  Fig. \ref{fig1}b. 

\begin{figure}[ht]
\begin{center}
\epsfxsize=0.5\textwidth
\epsfbox{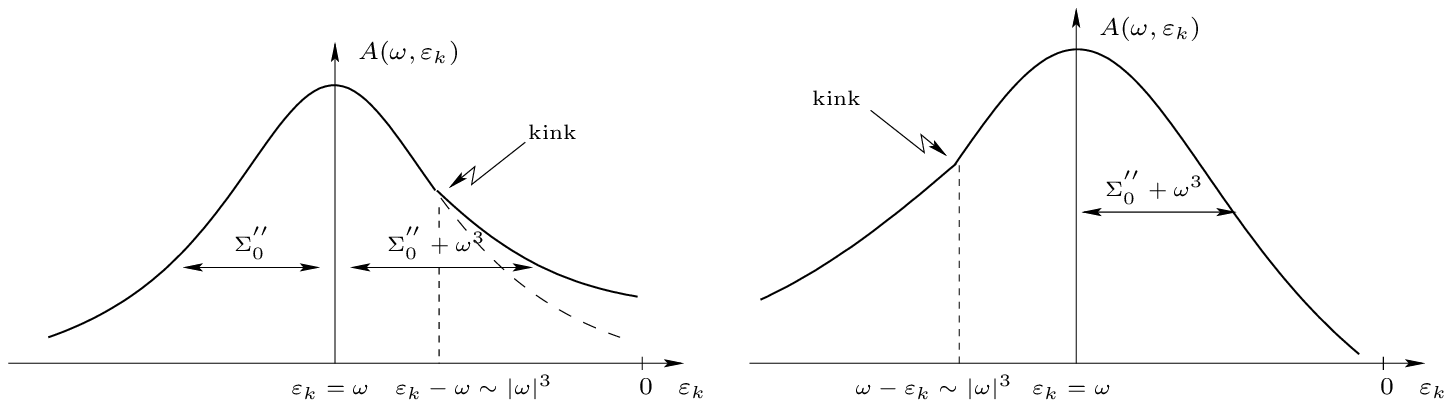}
\end{center}
\caption{ The behavior of the MDC spectral function 
 $A(k, \omega)$ as a function of $k$ at fixed $\omega$ 
for the interaction in the charge channel (a) and the spin channel (b). Observe that the kink in the spectral function is located on different sides of the mass shell in the two cases.}
\label{fig2}
\end{figure}

The different forms of $\chi_c$ and $\chi_s$ lead to different results for the
 full self-energy. We illustrate this for the $T=0$ case. 
For charge-mediated interaction, the softening of the square-root singularity in $\chi_s$ just implies that $Im \Sigma (\omega, k)$ becomes continuous. However, it still vanishes at $0>\omega > \epsilon_k$. We found
\begin{equation}
Im \Sigma (\omega, k) \propto |\omega|^3 \theta (\epsilon_k - \omega)
~ f\left(\frac{c^2 u_c^2 |\omega|^3}{(\epsilon_k - \omega) E^2_F}\right),
\end{equation}
 where $c \sim v_F/v_\Delta$, and $f(x)$ subject to $f(0) =1$ 
decreases with increasing $x$ and scales as $1/x$ at large $x$. 
This $1/x$ behavior
implies that very near the mass shell, 
 $Im \Sigma (\omega, E_k) \propto (\epsilon_k - \omega) \theta (\epsilon_k - \omega)$. The behavior of the spectral 
 function for this case is shown in Fig. \ref{fig2}a.
 For experimental comparison, we added a constant impurity 
scattering~\cite{peter}. 
The spectral function is anisotropic with respect to $\omega - \epsilon_k$,
 and  has a kink at $\omega -\epsilon_k \sim u^2_c \omega^3/E^2_F <0$, 
when the renormalization of the susceptibility becomes relevant. 

For a spin-mediated interaction, the situation is different. 
Due to the presence of the zero-sound pole in $\chi_s (\Omega, q)$, 
 $Im \Sigma (\omega, E_k)$ is non-zero at the mass shell, 
 and  only vanishes at 
$\omega - \epsilon_k > u^2_s |\omega|^3/E^2_F >0$,
 when the Cherenkov-type absorption becomes impossible.
 The spectral function now has the form of Fig. \ref{fig2}b.
 It again has a kink (more precisely, 
a square-root non-analyticity), but now the kink
 is located at $\omega - \epsilon_k >0$, 
i.e., on the other side of the mass shell. 
 We see therefore that interactions in the charge or in the spin channel 
lead to {\it qualitatively} different results for the spectral function. 

\begin{figure}[ht]
\begin{center}
\epsfxsize=0.33\textwidth
\epsfbox{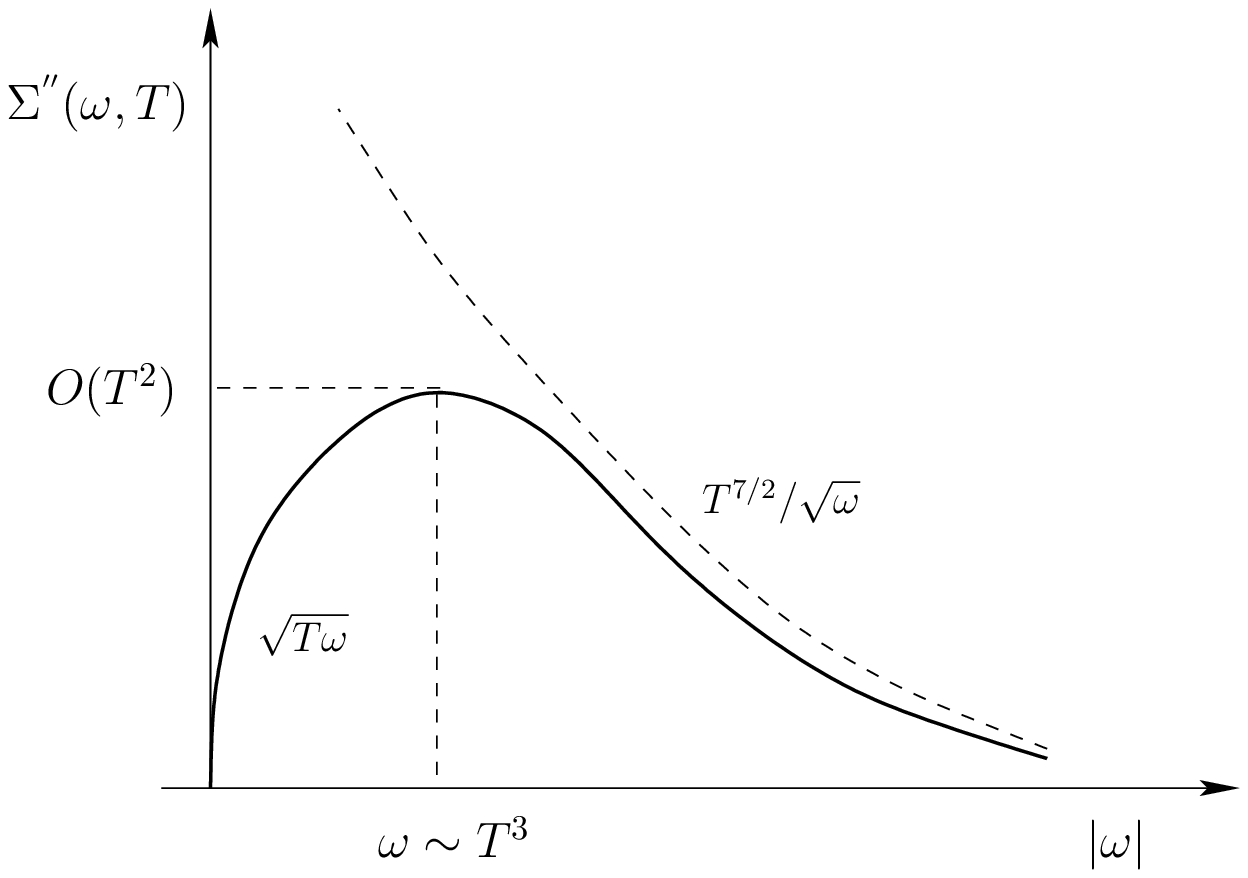}
\end{center}
\caption{The schematic behavior of Im $\Sigma (\omega, k_F, T)$ as a function of $\omega$ at a given $T$. The dashed line is the result of the second-order in perturbation. }
\label{fig3}
\end{figure}

A similar situation holds at a finite $T$. For breavity, we 
list the results for a 
 charge-mediated interaction, and set $k = k_F$.  For the most interesting case $T \gg |\omega|$, we found that $T^{7/2}/\sqrt{|\omega|}$ behavior holds 
down to   $ |\omega| \sim \omega_0 = u^2_c T^3/E^2_F$.
 Below this scale, $T^{7/2}/\sqrt{|\omega|}$
 behavior  crosses over to $\sqrt{T |\omega|}$. The crossover 
  behavior is somewhat involved, but the two limiting forms of $Im  \Sigma$ are
captured by a simple extrapolation formula
\begin{equation}
Im  \Sigma (\omega, T) \sim u_c ~\frac{T^2}{E_F}~ 
\Psi \left(\frac{|\omega|}{\omega_0}\right),~ \Psi (x) = \frac{\sqrt{x}}{x + 1}. \label{1}
\end{equation} 
We display this behavior in Fig.\ref{fig3}.  Note that 
 at the maximum, $Im \Sigma (\omega, T)$ scales as $T^2$. This is  
 qualitatively 
 different behavior from $Im \Sigma \propto T^3$, 
which would be the case if the mass shell
 singularity didn't exist. 

The unusual behavior of the spectral function has a profound influence  on the fermionic density of states $N(\omega) \sim \int d^2 x Im G (\omega, x)$. In particular,
$N(0) \propto T^{7/3}$.  It, however, makes little difference for  the tunneling density of states $N_{tunn} (\omega) \propto (1/T) \int  d\omega^\prime 
N(\omega^\prime)/\cosh^2 {(\omega + \omega^\prime)/2T}$. 
The latter one is non-zero already in the non-interacting $d-$wave gas, where
 $N(\omega) \propto |\omega|$. Then $N_{tunn} (0) \propto T$. 
The fermionic self-energy accounts for the corrections to the linear-$T$ behavior. Typical frequencies and $\epsilon_k$ 
in the integral for $N_{tunn} (0)$ are of the order of $T$, hence  relevant $Im \Sigma (\omega, k, T)$ are of the order of $T^{7/2}/\sqrt{|\omega|} \sim T^3$. 
As a result, $N_{tunn}/T = A + B T^2$ is analytic in $T$ 

To summarize, in this communication we re-considered 
the self-energy of a nodal fermion in a 2D $d-$wave 
 superconductor. We found that
 the $Im \Sigma (\omega, \epsilon_k, T)$ 
 is actually a complex function of $\omega, T$ and the
 quasiparticle energy $\epsilon_k$, and is qualitatively different 
from $Im \Sigma (\omega, T) \propto ~ {\text max}(\omega^3, T^3)$, suggested on general grounds. We found that the perturbative 
 self-energy is non-analytic near the mass shell at $T=0$. At a finite $T$, 
 it diverges as $T^{7/2}/\sqrt{|\omega|}$, when $\omega$ tends to zero, and 
 diverges even stronger near the mass shell.  
We demonstrated that the
 divergences in the self-energy for Dirac fermions
 originate from the same physics  as the mass shell singularity 
 in a 2D Landau Fermi liquid. 
We further demonstrated that  
the full summation of infinite diagrammatic series eliminates the divergences 
 and makes  $Im \Sigma (\omega, \epsilon_k, T)$ finite, 
 even at $\omega = \epsilon_k$. However,
 the full $\Sigma (\omega, \epsilon_k, T)$ at a given $T$ is non-monotonic as a function of $\omega$,  
and at a maximum is of order $T^2$ rather than $T^3$.  

We also found that the full spectral function has a kink, whose location compared to the mass shell depends on whether the dominant fermion-fermion 
interaction is in the spin or in charge channel. For negative $\omega$, relevant to ARPES experiments, kink is located at $\omega - \epsilon_k >0$ for 
 spin-mediated interaction, and at $\omega - \epsilon_k <0$ for charge-mediated interaction. We hope that careful ARPES studies of 
 the nodal spectral function in high $T_c$ materials below $T_c$ will be able to distinguish between charge and spin-mediated interactions.  
   
The doping dependence of the kink strength may also provide the information about the interaction.  Singular self-energy along the nodal direction is
 determined by intra-nodal scattering, and the magnitude of the effect 
 is determined by $U(q=0)$. This interaction is not enhanced if lowering the doping drives the system  closer
 to an instability at a finite $q$ ( e.g., an antiferromagnetic 
instability). On the other hand, if the system evolution towards half-filling
 is governed by 
low-energy fluctuations at small $q$, one should expect a strong enhancement of the kink strength at smaller doping. 

We thank  L. Glazman, P. Hirshfeld, 
 P. Johnson, D. Khveshchenko, D. Maslov and D. Scalapino 
 for useful discussions. 
 AVC acknowledges the support from Theory Institute for Strongly Correlated and Complex Systems at BNL, and NSF DMR 0240238. 
 AMT  acknowledges the support from
 US DOE under the contract  DE-AC02 -98 CH 10886.

\end{document}